\documentclass[a4paper,12pt]{article}
\usepackage[utf8]{inputenc}
\usepackage{natbib}
\usepackage{graphicx}

\title{Astrometric observations of outer Jovian satellites with the `Saturn' telescope. First results.}
\author{M.Yu. Khovritchev$^1$, A.P. Ershova$^1$, I.A. Balyaev$^1$,\\ D.A. Bikulova$^1$, I.S. Izmailov$^1$,  E.A. Roshchina$^1$,\\  V.V.Petjur$^1$, A.A. Shumilov$^1$, L.A. Maksimova$^1$,\\ K.I. Oskina$^1$, A.A. Apetyan$^1$, A.M. Kulikova$^1$  \\\\
{\small $^1$ Pulkovo Observatory, 65/1 Pulkovskoye chaussee, Saint Petersburg, 196140, Russia\footnote{e-mail: deimos@gao.spb.ru}}}

\begin{document} 

\maketitle

\begin{abstract}
The one-meter telescope-reflector `Saturn' (D=1~m, F = 4~m) was partially renovated at the Pulkovo observatory at the end of 2014. The telescope was equipped by CCD camera S2C with 14$\times$14 arcmin field of view and 824 mas per pix scale. The observations of outer Jovian satellites have been performed in a test mode since January 2015. The exposure time of 30 seconds allows us to obtain images of stars up to magnitude $19.5^m$ with the present state of the mirror and the equipment. The observations of outer Jovian satellites have been performed during testing period. These objects are interesting targets because their astrometric observations required to improve ephemeris and dynamic studies. Satellites positions have been determined on the basis of CCD images obtained within 6 nights. Astrometric reduction is performed by linear method using HCRF/UCAC4 and HCRF/URAT1. Internal accuracy of satellites positions has been estimated as 20 - 100~mas. The absolute values of residuals O-C do not exceed 100~mas in most cases. The independent tests have been carried out by the direct comparison with the results of observations of the Jovian satellite Himalia performed simultaneously by the Normal astrograph (the largest difference was 113~mas). This work has been partially supported by RFBR (12-02-00675-a) and the 22 Program of RAS Praesidium.

\end{abstract}

\section{Introduction}

Matters relating to the cosmogony and dynamics of the distant outer satellites families of giant planets occupy an important place among the fundamental problems of the Solar System research. Progress in solving these problems is impossible without high-precision astrometric observations of these bodies.

Even a significant amount of already accumulated ground-based CCD observations, ongoing and forthcoming observations in the framework of the GAIA mission is not sufficient to solve all the problems associated with the construction of ephemeris and understanding of the satellite dynamics~(\cite{Gomes_Junior2015}).

Ground-based observations of these satellites have an accuracy approximately 100 times less then the expected results of GAIA, but they can cover larger time intervals with higher density. Therefore both types of observations are carried out: ground-based (\cite{Gomes_Junior2015}) and the space observations of distant satellites of Jupiter~(\cite{Grav2015}).

First stage of works on restoration and adaptation of the one meter reflector `Saturn' for astrometric observations was completed at the end of 2014 at the Pulkovo Observatory. The first trial observations have been carried out since January 2015. This work allowed us to fully test the telescope and also to obtain some significant scientific data. In this paper we discuss the first results of astrometric observations of Jupiter distant satellites (Himalia, Elara, Pasiphae and Carme). 

Considered distant satellites of Jupiter belong to two families: Himalia (Himalia, Elara) and Pasiphae (Pasiphae). The first group has an average  jovicenter distance of about 11 million km, average inclination to the equatorial plane of the planet of about 30 degrees and the prograde motion. The Pasiphae group has a retrograde motion, average jovicenter distance of 23-28 million km and inclination of 147-152 degrees. 

Distant satellites location in the outer area of Jupiter satellites system indicates a possibility of capture of these bodies from interplanetary space. They have not undergone essential changes during the period of time comparable to the age of the Solar system (4.5 billion years), and, therefore, they may keep the information about the early stages of the Solar system formation.

Previously, data about the Jovian outer irregular satellites motion has been used to determine the mass of Jupiter~(\cite{Bykova1979}). At present the theories of satellites motion are calculated basically using the analytical models (\cite{Emelyanov2005}). The gravitational influence of Jupiter, its Galilean satellites, Saturn and the Sun are taken into account at the motion theories (\cite{Jacobson2000}).

Regular observations of Jupiter distant satellites (as well as other distant satellites of giant planets) are carried out at the  ESO (Chile), Pico los Dias (Brazil) observatories, and at the observatory of Haute Provence (France)~(\cite{Gomes_Junior2015}). Accuracies of coordinates are about  60~-~80~mas and depend on the object brightness. The problem of distant planets irregular satellites orbits uncertainty is also being studied (\cite{Jacobson_et_al2012}).

The next sections provide brief information about the telescope `Saturn', describe the methods of observation and image processing, the results of determining equatorial coordinates of the satellites.

\section{Observations and image processing}\label{obs}

\begin{figure}[h]
    \begin{center}
       \includegraphics[width=\columnwidth]{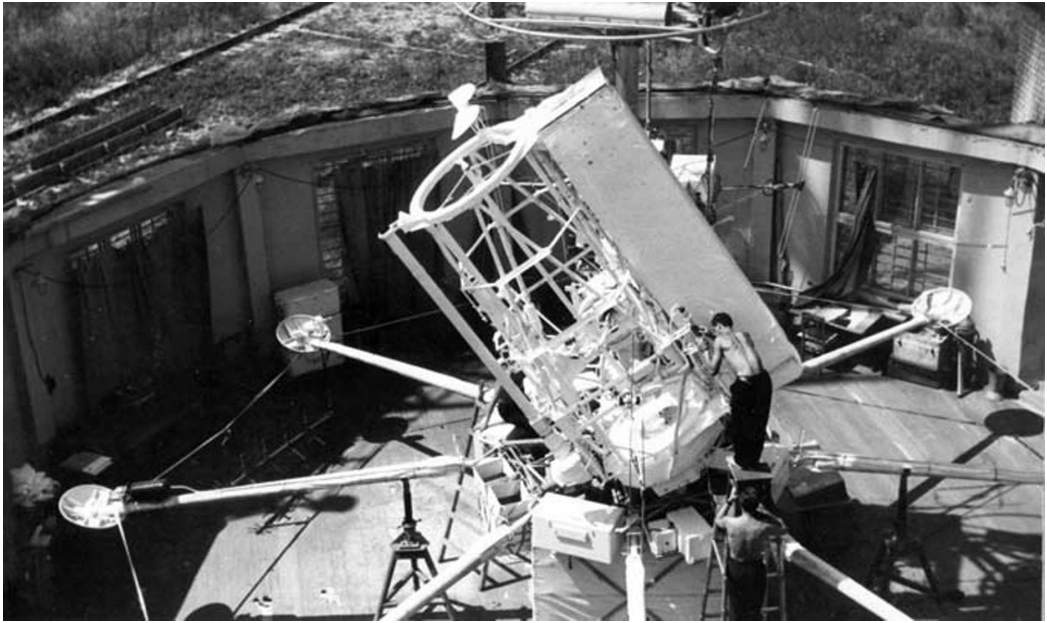}
       \caption{The one-meter telescope-reflector `Saturn' in 1970s.} 
         \label{Fig1}
    \end{center}
\end{figure}

The ex-stratospheric solar telescope `Saturn' was restored and adapted to the astrometric observations in the Laboratory of astrometry and stellar astronomy of the Pulkovo Observatory at the end of 2014. This telescope was constructed at the Kazan Optical and Mechanical Plant in the early 1970s for the fourth flight of the Soviet stratospheric solar station `Saturn', which was raised for the observations to the 20~km altitude by an aerostat (\cite{Krat_Kotlyar1976}). General view of the telescope is seen in Fig.~\ref{Fig1}. Originally, the telescope had a focal length equivalent to 120 m. Unprecedentedly high resolution images (0.12~arcsec) of solar granulation and spots had been obtained  with this telescope  during the flight in 1973 (\cite{Krat_et_al1974}).

The telescope was stored at the plant after the solar station fulfilled its task and the project ended. Later in 1988-1996 the telescope was installed at Pulkovo Observatory, and It was modernized into the ground-based solar telescope under the guidance of L.D.~Parfinenko (\cite{Parfinenko2011}). The test observations of the Sun were carried out within two years and showed that using this telescope it is impossible to obtain Solar images of perfect quality with the conditions of the local astroclimate. It was proposed to modernize the instrument for nighttime astrometric observations. Up to 2013 the telescope was in the storage.

Over the last decade the astrometric research development has a tendency to study more and more faint objects. Therefore, the use of a telescope with a larger aperture (larger than our available telescopes) is very important for our astrometric purposes. The large volume of works on the telescope and the pavilion restoration was performed during 2013 - 2014. The telescope parabolic mirror has a diameter of 1~m with a focal length of 4~m. The secondary mirror was removed and the CCD camera S2C was installed in the direct focus of the primary mirror. Temporarily, observations are carried out without filters. The CCD camera specifications are: FOV – 14$\times$14~arcmin, scale – 824~mas/px, size - 1024$\times$1024 px, pixel size is 16~mkm. The telescope is mounted on the alt-azimuth mount and is driven by stepper motors. The Canon camera with a large field of view and a long focus lens (F = 300~mm) is used for positioning and guiding. The frames are automatically recognized by the specially developed software. This gives the possibility to use the auto guiding mode.

\begin{figure}[h]
    \begin{center}
       \includegraphics[width=0.4\columnwidth]{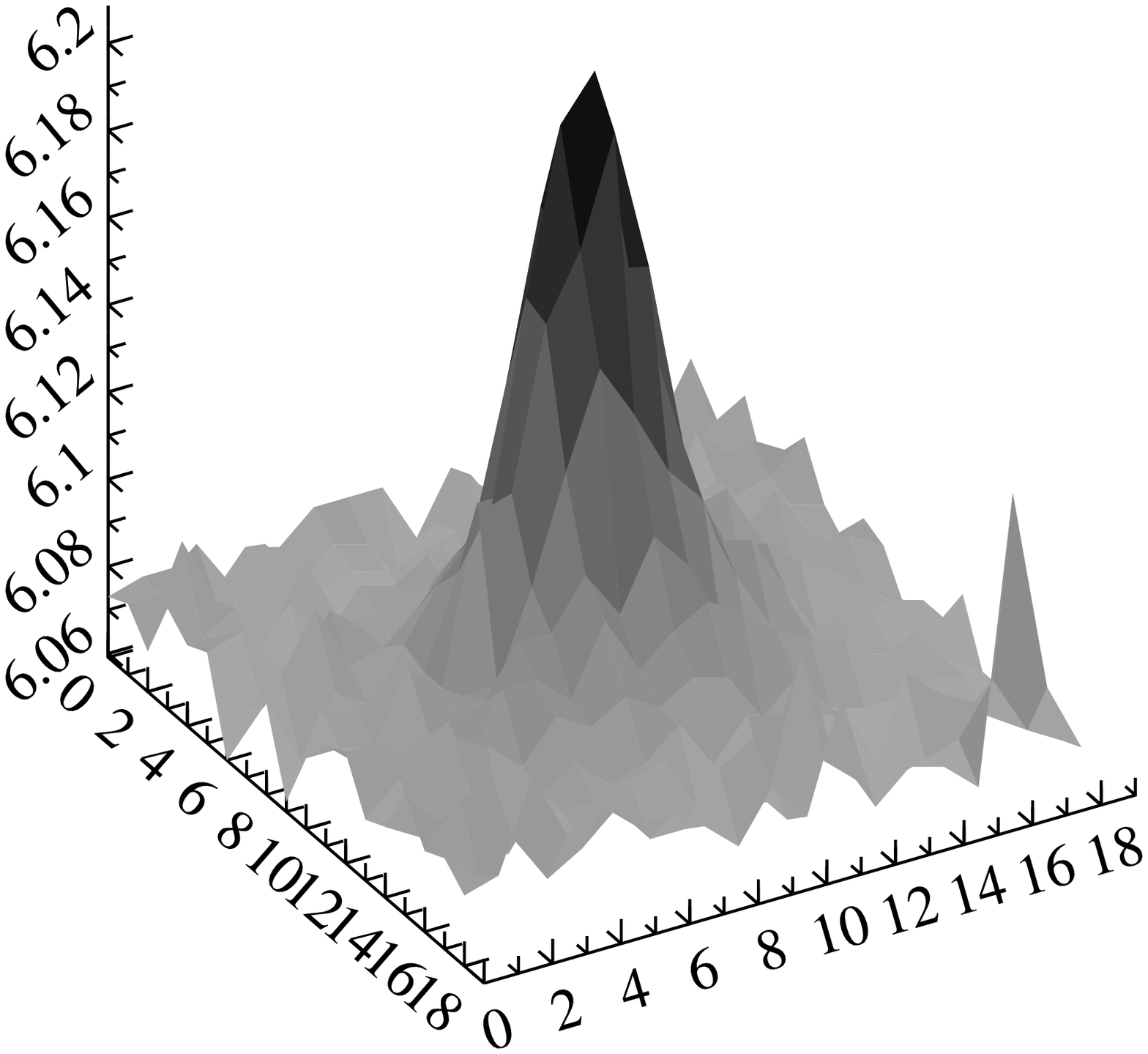}
       \includegraphics[width=0.4\columnwidth]{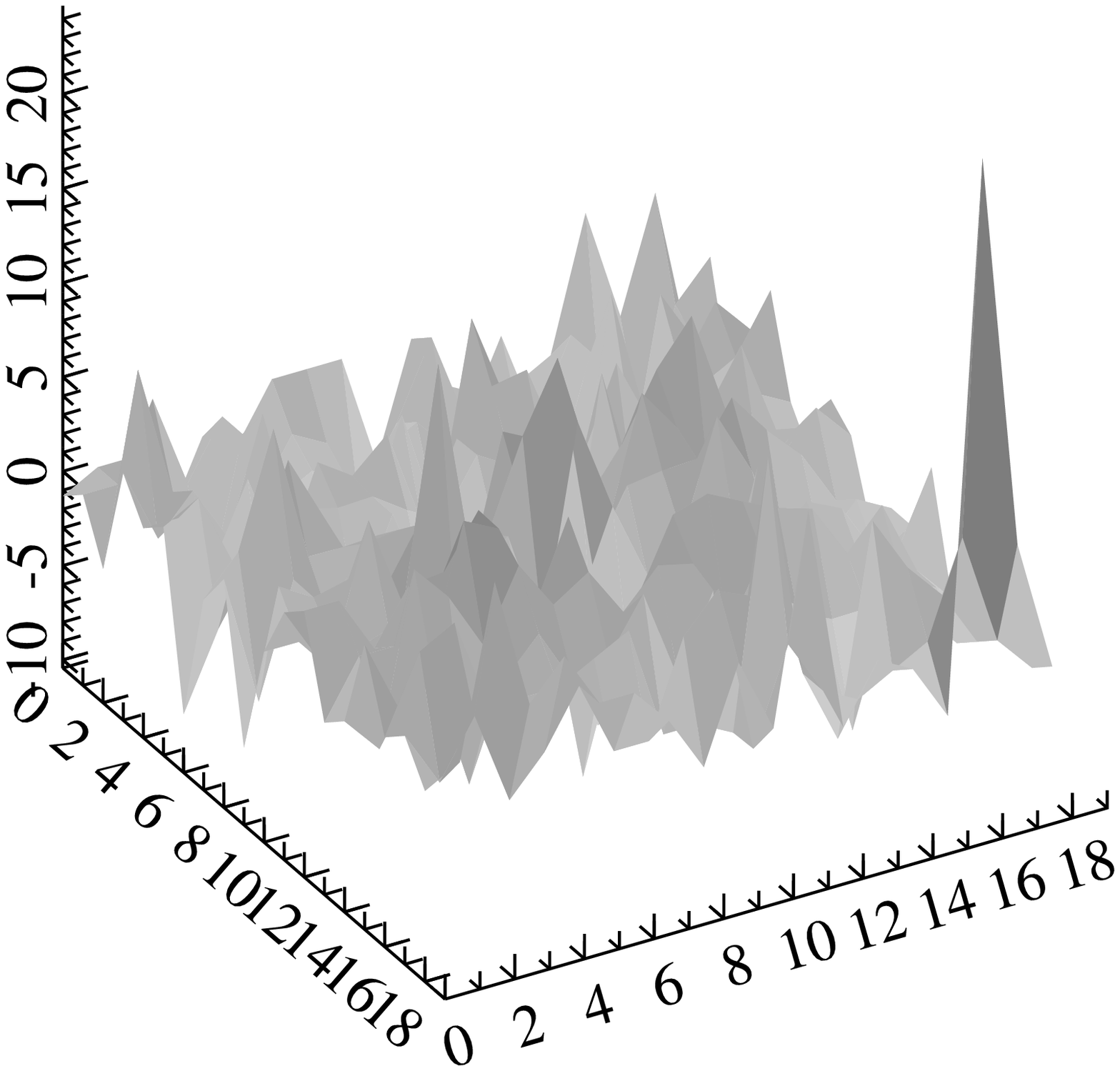}
       \caption{Example of the Pasiphae image that was built by summing of four ccd-frames. To the left - the original image (countings are shown in logarithmic scale), to the right - the result of subtracting a model of an image obtained by using shapelet-decomposition of the original image. The signal-to-noise ratio is about 15 for the maximum counting of the image.} 
         \label{Fig2}
    \end{center}
\end{figure}

The first experimental observations were carried out in late 2014. Test observations for various research programs were completed in May 2015 and it was concluded that the telescope is suitable for astrometric observations. Jovian satellites Himalia, Elara, Pasiphae and Carme were chosen for the first observations. They are bright enough and located relatively far from Jupiter, so Jupiter’s halo does not interact with images of these satellites. The observations were made near meridian with hour angles of less than 2~h. Debugging and testing of the various components of the instrument was carried out at the same time with the satellites observations. The observations within the Pulkovo program of the large proper motion stars research and the observations of selected asteroids were also performed. 

Six observational nights were intended for distant satellites of Jupiter. However, this is enough to make a conclusion about advisability of further astrometric observations with `Saturn' telescope. The data was obtained by the series of ccd-frames (10-20 frames) with exposure times of 10, 20  and 30 sec. The parameters of WCS that are necessary for further analysis were recorded into FITS-files headers. Summing of CCD-frames was used to improve the signal-to-noise ratio for the faint satellites, in dependence of the image quality the series of 4 - 6 frames were summed. The linear model parameters of transformation from the current frame to the standard one were calculated using bright reference stars before the procedure of summation. The shapelet decomposition (\cite{Massey_Refregier2005}) was used to determine the parameters of images of satellites and reference stars in each frame. 

\begin{figure}[h]
    \begin{center}
       \includegraphics[angle=-90, width=0.45\columnwidth]{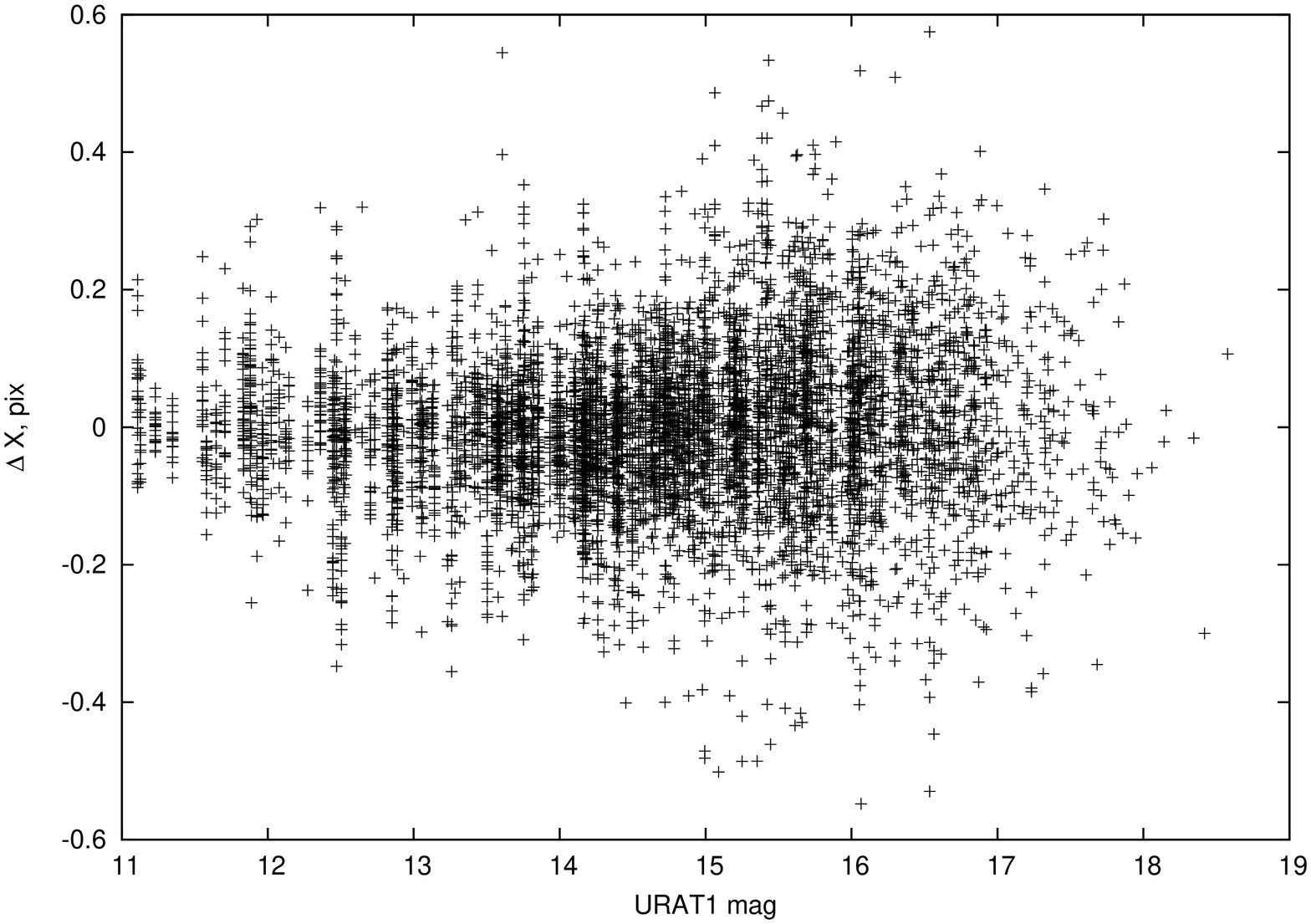}
       \includegraphics[angle=-90, width=0.45\columnwidth]{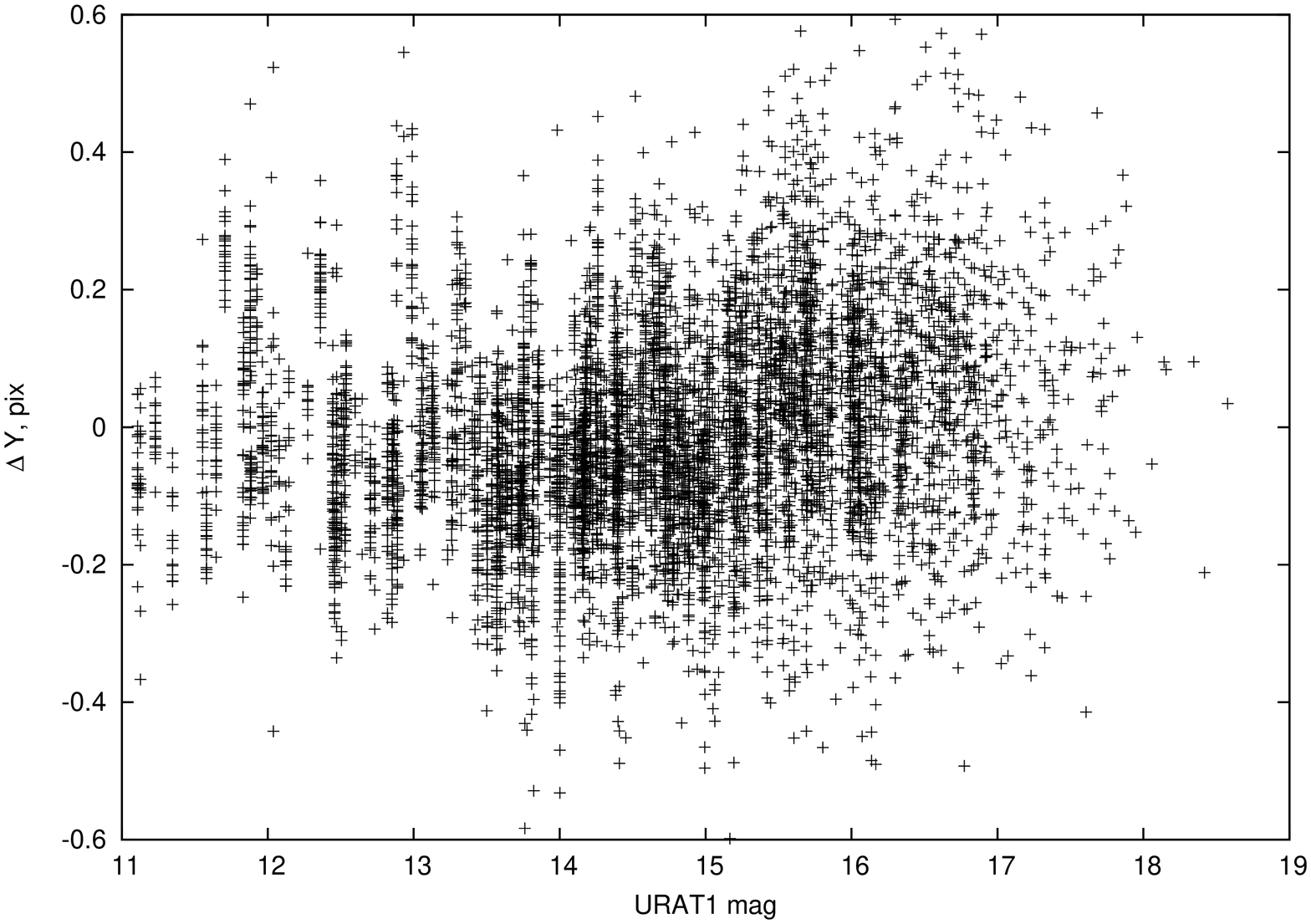}
       \caption{The residuals of reference star pixel coordinates as function of magnitude. There is a noticeable trend for y-coordinate.} 
         \label{Fig3}
    \end{center}
\end{figure}

As a result, the pixel coordinates of the targets and stars photocenters have been determined. To illustrate the method there is a diagram in Fig.~\ref{Fig2} that shows the structure of the Pasiphae image and the result of subtracting the model image built on the basis of shapelet-decomposition parameters. Initially the astrometric reduction of ccd-frames was made with IZMCCD package (\cite{Izmailov_et_al2010}\footnote{http://izmccd.puldb.ru}). The UCAC4 catalogue (\cite{Zacharias_et_al2013}) was used as a reference one. It turned out that even without a rigorous analysis of systematic errors of stellar coordinates the standard deviation of both calculated coordinates is typically less than 150 - 200~mas, and value of `O-C' residuals for satellites coordinates usually does not exceed 200 mas with good internal convergence (of the order of 100 - 150~mas) for almost all series. However the UCAC4 catalogue has relatively low accuracy of coordinates and proper motions for the relatively faint stars of $15^m~-~16^m$. Therefore, the URAT1 catalogue (\cite{Zacharias_et_al2015}), containing positions of the stars up to $19^m$ at the epoch of 2013, was used for the final processing. This made it possible to correctly take into account the errors of faint reference stars positions and to reduce the dependence of results on proper motion errors. Fig.~\ref{Fig3} shows the dependence of the pixel coordinates of reference stars residuals on magnitude. As it follows from Fig.~\ref{Fig3},  the accuracy of measuring is higher for objects that are up to $16^m$ (residuals are within $\pm$0.25~pix). Measurement accuracy is significantly lower for the fainter objects, there is a residual's dependence on the brightness, it is most noticeable along the axis Y, so it is necessary to more accurately recognize and correct the systematic errors of positions of reference stars and targets.

\begin{figure}[h]
    \begin{center}
       \includegraphics[width=0.45\columnwidth]{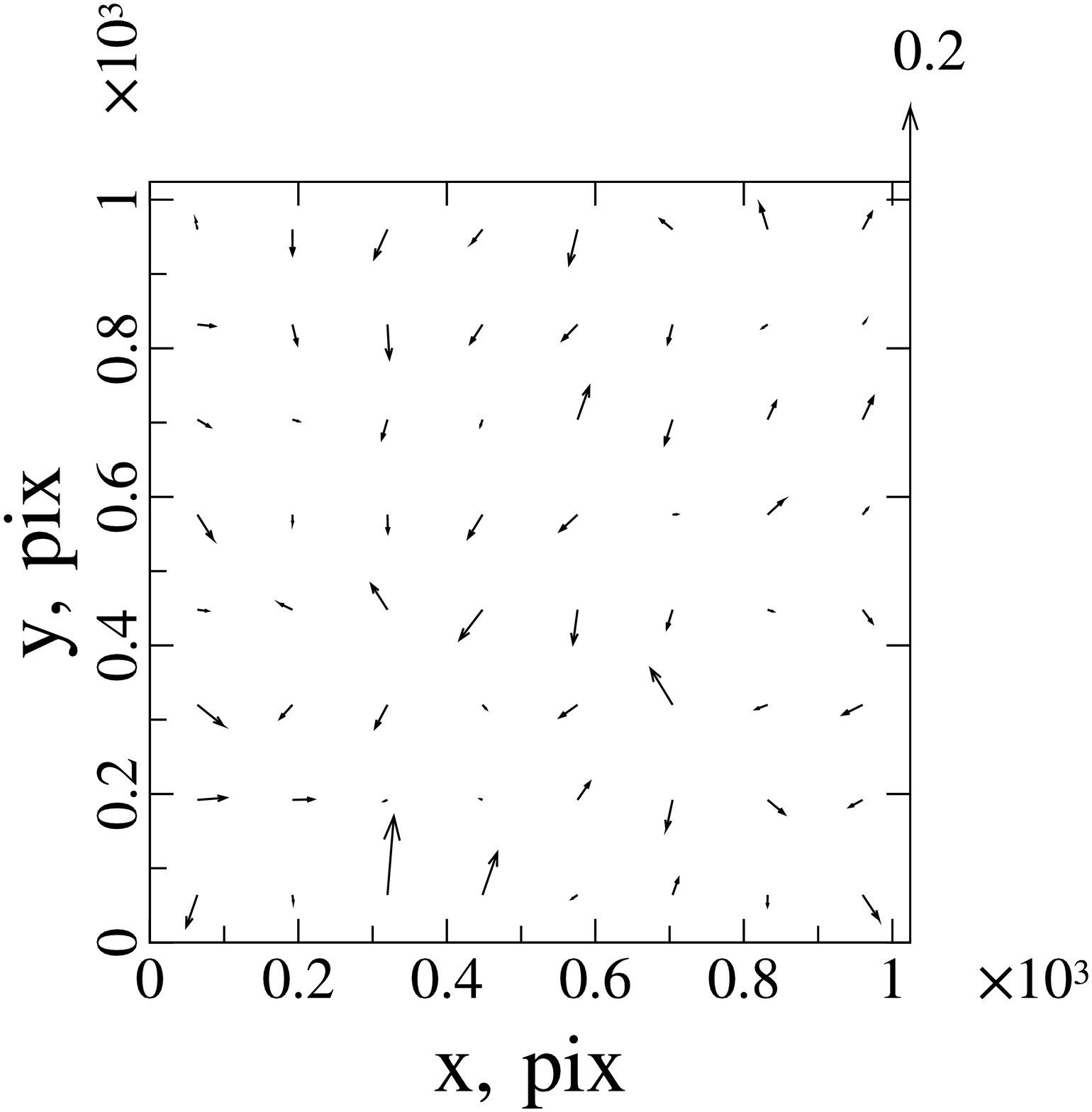}
       \includegraphics[width=0.45\columnwidth]{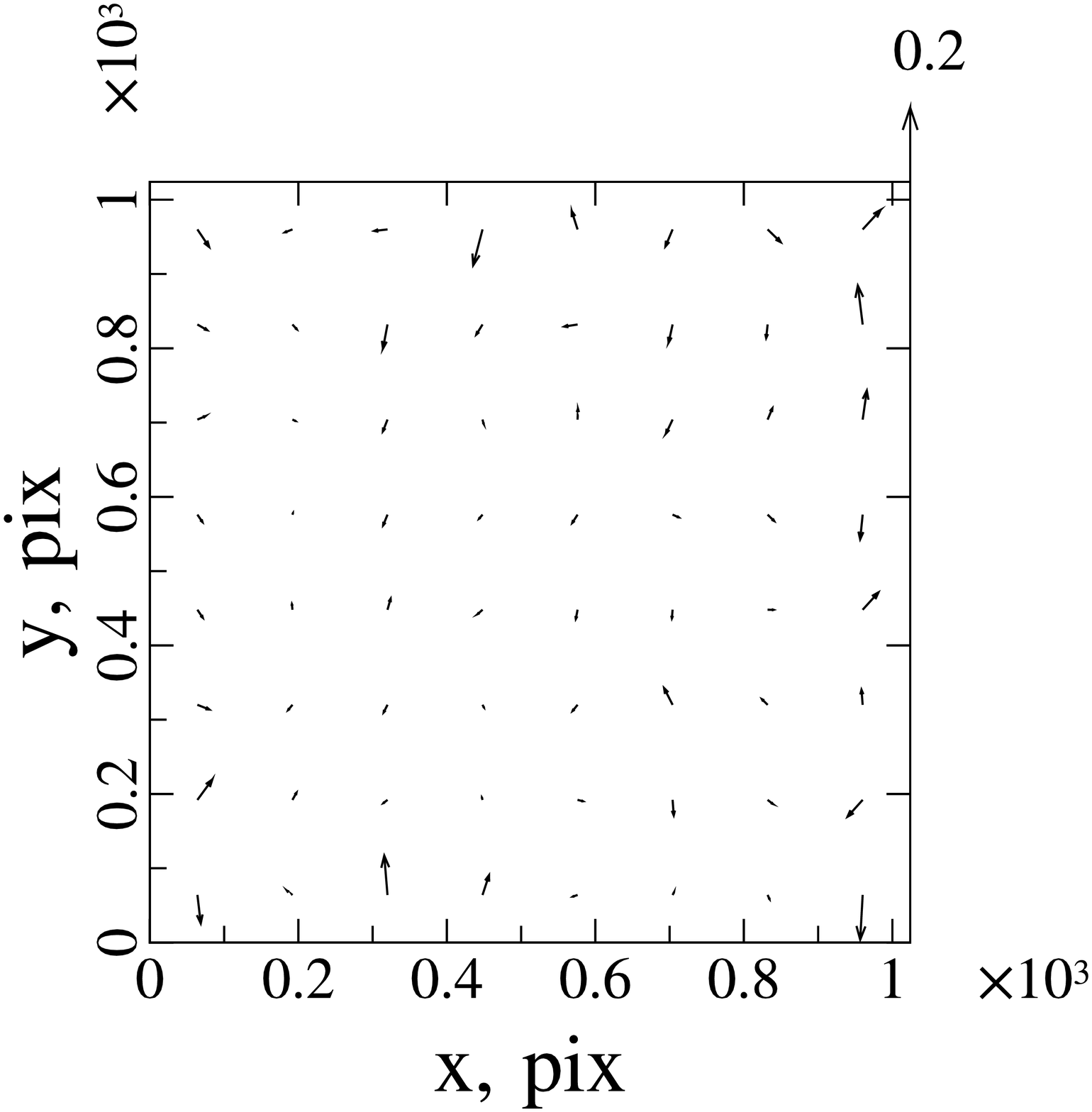}
       \caption{Vector fields of reference star's residuals in the range of $14^m$ - $18^m$ before (to the left) and after correction (to the right). The scale vector is shown in the upper right corner. These fields have been built on the different sets of residuals obtained with reductions of various series of CCD frames.} 
         \label{Fig4}
    \end{center}
\end{figure}

Vector fields of reference stars pixel coordinates residuals have been built for three ranges of magnitudes (from $12^m$ to $18^m$ in increments of $2^m$) and for the whole FOV. Averaging of the residuals has been carried out within square areas (side of 128~pix), each containing of 20 - 60 reference stars. Total amount of considered reference stars and their residuals is more then 9200. Corresponding corrections that depend on the pixel positions and star magnitudes were applied to the pixel coordinates of reference stars and satellites before the final reduction. The reduction has been made by linear method using six constants. Fig.~\ref{Fig4} demonstrates the relevance of this approach to systematic errors correction. Number of reference stars for the different observational series varied between 10 to 60. The errors of determining of the coordinates ranged from 40 to 200~mas.

\section{Results and their analysis}\label{res}

Positions of selected distant satellites of Jupiter included into the program of the first observations, their differences with ephemeris (O-C) and the errors are shown in Table~\ref{Tab1}. The planetary theory EPM-2011 has been used for ephemeris. Satellite ephemerides have been provided with  MULTI-SAT service (\cite{Emelyanov_Arlot2008}\footnote{http://lnfm1.sai.msu.ru/neb/nss/nssephmr.htm}). Table~\ref{Tab1} shows that the internal accuracy of mean positions of the Himalia is better than 50 mas. For fainter satellites the accuracy is somewhat lower, but in almost all cases it is better than 100~mas. Values (O-C) also do not exceed 100~mas in most cases.

\begin{table}
 \caption{Equatorial coordinates of Himalia, Elara, Pasiphae and Carme, their standard errors and O-C differences.}\label{Tab1}
\vspace{5mm}
\begin{tabular}{rrrrrrrr}

Date (UTC) & RA&Dec& \multicolumn{2}{c}{(O-C)}& $\varepsilon_\alpha$&$\varepsilon_\delta$\\
               & \multicolumn{2}{c}{J2000} & RA & Dec&  &\\
YYYYMMDDhhmmss & h m s& $^\circ~ '~ ''$ & \multicolumn{2}{c}{mas}& mas &mas\\\hline
&\multicolumn{6}{c}{J6 Himalia, $15.1^m$}\\\hline
2015 02 10 22 25 44.4& 09 18 12.760& 17 09 59.62& -172& 164&  13&  9\\
2015 02 15 23 27 03.2& 09 16 10.086& 17 20 38.70&   13&  93&  29& 41\\
2015 02 17 22 04 50.9& 09 15 23.921& 17 24 27.56&  -16&   7&  20& 34\\
2015 03 12 20 56 42.5& 09 07 46.992& 17 55 11.97&  -10& 139&  16& 20\\
2015 03 16 20 54 04.4& 09 06 50.558& 17 57 39.57&   -4&  88&  18& 13\\\hline
&\multicolumn{6}{c}{J7 Elara, $16.9^m$}\\\hline
2015 02 15 22 25 22.2& 09 13 58.359& 16 37 09.13&    6& 217&  43& 23\\
2015 02 17 23 12 53.6& 09 12 42.491& 16 42 05.66&  -68&  79& 173& 95\\
2015 03 12 21 12 57.8& 09 01 07.329& 17 30 13.79&  125&  97&  81& 57\\
2015 03 17 21 08 41.2& 08 59 27.190& 17 37 59.84&   54& -84&  70& 51\\\hline
&\multicolumn{6}{c}{J8 Pasiphae, $17.3^m$}\\\hline
2015 02 15 23 14 28.7& 09 07 43.027& 17 13 18.38&   1& -53&  52& 87\\
2015 03 12 20 42 42.2& 08 57 09.326& 18 10 35.44&  23&  13&  64& 57\\
2015 03 16 20 08 42.5& 08 56 01.986& 18 16 57.73& 228&  97&  65& 78\\\hline
&\multicolumn{6}{c}{J11 Carme, $18.2^m$}\\\hline
2015 03 16 21 16 23.8& 09 00 03.200& 17 51 18.19& -293& 63& 79& 153\\\hline
\end{tabular}
\end{table}

In addition to the observations with the `Saturn' telescope the observations of Himalia have been carried out with the Normal Astrograph  of the Pulkovo Observatory (D=330~mm, F=3467~mm). The observations with the Normal Astrograph are carried out using SBIG ST-L-11K CCD camera (FOV 35$\times$23~arcmin, scale 533~mas/pix). Data processing method is equivalent to that described in the previous section. The astrometric reduction has been carried out by the six constant method with the URAT1 catalogue as a reference. Errors of one position (standard deviations) have been estimated as 50 - 100~mas. The systematic errors correction of stars coordinates has been carried out according to the method described in the previous section. The observations were made with both telescopes during two nights on February 15$^{th}$ and 17$^{th}$  at the close time moments. In both cases, velocity components of satellite motions on the celestial sphere have been calculated using the observations with the `Saturn' telescope. 

\begin{table}
 \caption{Equatorial coordinates of Himalia obtained with the Normal Astrograph.}\label{Tab2}
\vspace{5mm}
\begin{tabular}{rrrrrrrr}

Date (UTC) & RA&Dec& \multicolumn{2}{c}{(O-C)}& $\varepsilon_\alpha$&$\varepsilon_\delta$\\
               & \multicolumn{2}{c}{J2000} & RA & Dec&  &\\
YYYYMMDDhhmmss & h m s& $^\circ~ '~ ''$ & \multicolumn{2}{c}{mas}& mas &mas\\\hline

2015 01 21 23 59 18.0& 09 26 22.964& 16 20 04.46& 292&  22& 105&  48\\
2015 02 15 22 46 12.0& 09 16 10.777& 17 20 35.18&   1& -30&  28&  73\\
2015 02 17 22 09 37.0& 09 15 23.842& 17 24 28.02&  -4&  81&  79&  41\\
2015 03 09 22 53 40.0& 09 08 33.238& 17 52 50.62&   7& -37& 117& 186\\\hline

\end{tabular}
\end{table}

It has allowed us to interpolate the positions of Himaia at the mean time moments of Normal Astrograph observations. The results are shown in Table~\ref{Tab2}. This approach has been applied to perform the comparison without using the satellite motion theory. The corresponding differences between results of observations `Saturn - Normal Astrograph' are:

\vspace{5mm}
\begin{tabular}{r}
46~mas and 113~mas at 2015~02~15, 22h~46m~12s.0 (UTC),\\
–16~mas and –92~mas at 2015~02~17, 22h~09m~37s.0 (UTC).\\
\end{tabular}

\section{Conclusion}\label{concl}
In observational season of 2014-2015 at the Pulkovo Observatory the one meter reflector `Saturn' started to work as an astrometric telescope for the first time. The first experimental observations included the selected distant satellites of Jupiter - Himalia, Elara, Pasiphae and Carme. Analysis of the material of these observations has been made, it has allowed us to reveal and significantly reduce the influence of systematic errors. 

These systematic errors were initially small. It characterizes the `Saturn' telescope as an instrument for high accuracy astrometry of relatively faint objects of $12^m$ - $18^m$. 

In the future, the limiting magnitude that is available for telescope may be increased by realuminizing of the mirror and with equipment improvements.  And there are some prospects for improving the observations accuracy.

As a result, we have obtained five astrometric positions of Himalia, four positions of Elara, three - for Pasiphae and one - for the Carme. Convergence of `O-C' is within 20 - 100~mas in most cases. Differences with ephemeris `O-C' are within $\pm$100~mas, which is consistent with the typical values obtained during observations at the various telescopes of the world~(\cite{Gomes_Junior2015}).

Simultaneous observations of Himalia, that were performed with Normal Astrograph of the Pulkovo observatory, confirmed the good quality of astrometric observations with the `Saturn' one meter reflector. 

\section*{Acknowledgements}
The authors thank Dr. Y.A.~Nagovitsyn and Dr. L.D.~Parfinenko for supporting the project. This work has been carried out within the scope of projects of major planets satellites observations and has been partially supported by the grant RFBR12-02-00675-a and the Programs of the Presidium of RAS 22 and 9.


\begin{thebibliography}{99}

\bibitem[\protect\citeauthoryear{Parfinenko}{2011}]{Parfinenko2011} Parfinenko L.D. 2011. Solar atmosphere structures in various time and spatial scales. Doctoral thesis, Saint Petersburg, p. 46. (in Russian).
	
\bibitem[\protect\citeauthoryear{Krat and Kotlyar}{1976}]{Krat_Kotlyar1976} Krat V.A. and Kotlyar L.M. 1976. Stratospheric astronomy, Saint Petersburg, Science. (in Russian).

\bibitem[\protect\citeauthoryear{Krat}{2006}]{Krat2006} Krat T.V. 2006. About flies of stratospheric station, Astronomical Calendar, p. 175. (in Russian).

\bibitem[\protect\citeauthoryear{Bykova}{1979}]{Bykova1979} Bykova, L. E. 1979. Institut Teoreticheskoi Astronomii, Biulleten', 14, 7, p. 402.

\bibitem[\protect\citeauthoryear{Gomes-Junior et al.}{2015}]{Gomes_Junior2015}  Gomes-Junior A. R., Assafin M., Vieira-Martins R., Arlot J.-E., Camargo J. I. B., Braga-Ribas F., da Silva Neto D. N., Andrei A. H., Dias-Oliveira A., Morgado B. E., Benedetti-Rossi G., Duchemin Y., Desmars J., Lainey V., Thuillot W. 2015. arXiv:1506.00045

\bibitem[\protect\citeauthoryear{Grav et al.}{2015}]{Grav2015}  Grav T., Bauer J. M., Mainzer A. K., Masiero J. R., Nugent C. R., Cutri R. M., Sonnet S., Kramer E. 2015. arXiv:1505.07820.

\bibitem[\protect\citeauthoryear{Emel'yanov}{2005}]{Emelyanov2005} Emelyanov N. V. 2005. A\&A, 435, 3, 1173.

\bibitem[\protect\citeauthoryear{Emel'yanov and Arlot}{2008}]{Emelyanov_Arlot2008} Emel'yanov N. V. and Arlot J.-E. 2008. A\&A, 487, 759.

\bibitem[\protect\citeauthoryear{Jacobson}{2000}]{Jacobson2000} Jacobson R. A. 2000. AJ, 120, 5, 2679.

\bibitem[\protect\citeauthoryear{Jacobson et al.}{2012}]{Jacobson_et_al2012} Jacobson R., Brozovic M., Gladman B., Alexandersen M., Nicholson P. D., Veillet C. 2012. AJ, 144, 5, 132.

\bibitem[\protect\citeauthoryear{Izmailov et al.}{2010}]{Izmailov_et_al2010} Izmailov I.S., Khovricheva M. L., Khovrichev M. Yu., Kiyaeva O.V., Khrutskaya E.V., Romanenko L.G., Grosheva E.A., Maslennikov K.L., Kalinichenko O.A. 2010. Astronomy Letters, 36, 5, 349.

\bibitem[\protect\citeauthoryear{Krat et al.}{1974}]{Krat_et_al1974} Krat V. A., Dul'Kin L. Z., Validov M. A., Vakhrameev I. Ya., Karpinskij V. N., Muzalevskij Yu. S., Nikolaev R. P., Protsenko B. A., Sobolev V. M., Tabakova Z. N., Shakhbazyan Yu. L 1974. Astronomicheskii Tsirkulyar, 807, 1.

\bibitem[\protect\citeauthoryear{Massey and Refregier}{2005}]{Massey_Refregier2005} Massey R., Refregier A. 2005. MNRAS, 363, 1, 197.

\bibitem[\protect\citeauthoryear{Zacharias et al.}{2013}]{Zacharias_et_al2013} Zacharias N., Finch C.T., Girard T. M., Henden A., Bartlett J. L., Monet D. G., Zacharias M. I. 2013. AJ., 145,  2,  44.

\bibitem[\protect\citeauthoryear{Zacharias et al.}{2015}]{Zacharias_et_al2015} Zacharias N., Finch C.T., Subasavage J.P., Tilleman T., DiVittorio M., Harris H.C.,  Rafferty T., Wieder G., Ferguson E., Kilian C., Rhodes A., Schultheis M. 2015. American Astronomical Society, AAS Meeting 225, 433.01.


\end{thebibliography}
\end{document}